\documentclass[aps,prl,twocolumn,superscriptaddress,showpacs]{revtex4}
\usepackage{graphicx}

\newcommand{\gk}{\ensuremath{\mathrm{\thinspace GK}}}
\newcommand{\etal}{ \textit{et~al.}}
\newcommand{\eg}{e.g.}

\newcommand{\msun}{\ensuremath{M_{\odot}}}

\newcommand{\gcc}{\ensuremath{\mathrm{\thinspace g \thinspace cm^{-3}}}}

\begin{document}

\title{The consequences of nuclear electron capture in core collapse supernovae}


\author{W. R. Hix}
\author{O. E. B. Messer}
\affiliation{Department of Physics and Astronomy, University of Tennessee, Knoxville, TN 37996-1200 USA}
\affiliation{Physics Division, Oak Ridge National Laboratory, Oak Ridge, TN 37831-6354 USA} 
\affiliation{Joint Institute for Heavy Ion Research, Oak Ridge National Laboratory, Oak Ridge, Tennessee 37831-6374 USA}
\author{A. Mezzacappa}
\affiliation{Physics Division, Oak Ridge National Laboratory, Oak Ridge, TN 37831-6354 USA} 
\author{M. Liebend\"orfer}
\affiliation{Canadian Institute for Theoretical Astrophysics, Toronto ON M5S 3H8 Canada}
\affiliation{Department of Physics and Astronomy, University of Tennessee, Knoxville, TN 37996-1200 USA}
\affiliation{Physics Division, Oak Ridge National Laboratory, Oak Ridge, TN 37831-6354 USA} 
\author{J. Sampaio}
\author{K. Langanke}
\affiliation{Institute of Physics and Astronomy, University of \AA rhus, DK-8000 \AA rhus C, Denmark} 
\author{D. J. Dean}
\affiliation{Physics Division, Oak Ridge National Laboratory, Oak Ridge, TN 37831-6354 USA} 
\author{G. Mart\'inez-Pinedo}
\affiliation{Institut d'Estudis Espacials de Catalunya, E-08034 Barcelona, Spain}
\affiliation{Instituci— Catalana de Recerca i Estudis Avanats, E-08010 Barcelona, Spain}

\date{\today}

\begin{abstract}
The most important weak nuclear interaction to the dynamics of stellar core
collapse is electron capture, primarily on nuclei with masses larger
than 60.  In prior simulations of core collapse, electron capture on
these nuclei has been treated in a highly parameterized fashion, if not
ignored.  With realistic treatment of electron capture on heavy nuclei
come significant changes in the hydrodynamics of core collapse and
bounce. We discuss these as well as the ramifications for the
post-bounce evolution in core collapse supernovae.
\end{abstract}


\maketitle


Core collapse supernovae are among the most energetic events in the
universe, emitting $10^{46}$ J of energy, mostly in the form of neutrinos. 
These explosions mark the end of the life of a massive star, the formation
of a neutron star or black hole and play a preeminent role in the cosmic
origin of the elements. With the formation of an \emph{iron core} in a
massive star (containing the maximally-bound iron and neighboring nuclei),
thermonuclear energy is no longer available to slow the inexorable
contraction that results from a star's self gravity.  Once this cold iron
core grows too massive to be supported by the pressure of degenerate
electrons, core collapse ensues.  In the inner region of the core, this
collapse is subsonic and homologous, while the outer regions collapse
supersonically.  When the inner core exceeds nuclear densities, it
stiffens, halting the collapse.  Collision of the supersonically infalling
outer core with this stiffened inner core produces the \emph{bounce shock},
which initially drives outward the outer layers of the iron core.  However,
this bounce shock is sapped of energy by the escape of neutrinos and
nuclear dissociation and stalls before it can drive off the envelope of the
star (see, \eg, \cite{BuLa85,ThBP03}).  The intense neutrino flux, which is
carrying off the binding energy of the proto-neutron star (PNS), heats
matter between the neutrinospheres and the stalled shock. In the
\emph{neutrino reheating} paradigm, this heating reenergizes the shock,
which drives off the concentric layers of successively lighter elements
that lie above the iron core, producing the supernova.

Unfortunately, simulations exploring the neutrino reheating paradigm often
fail to produce explosions.  The failure of recent spherically symmetric
multigroup Boltzmann simulations \cite{RaJa00,MLMH01,LMTM01} to produce
explosions has removed incomplete neutrino transport as a potential cause
of this failure.  Models that break the assumption of spherical symmetry
have achieved some success, either by an increase in the neutrino
luminosity due to fluid instabilities within the proto-neutron star
\cite{WiMa93} or by enhancement of the efficiency of the neutrino heating
by large scale convection behind the shock \cite{HBHF94,BuHF95,FrWa02}. The
PNS instabilities are driven by lepton and entropy gradients, while
convection behind the shock originates from gradients in entropy that
result from the stalling of the shock and grow as the matter is heated from
below. However, even with such enhancements, explosions are not guaranteed
\cite{JaMu96,MCBB98,BRJK03}.  A third potential cause of the failure to
produce explosions in numerical models is incomplete or inaccurate
treatment of the wide variety of nuclear and weak interaction physics that
is important to the supernova mechanism.  Once the supernova shock forms,
emission and absorption of electron neutrinos and antineutrinos on the
dissociation-liberated free nucleons are the dominant processes. However,
during core collapse, electron capture on nuclei plays an important and, as
we will demonstrate, dominant role by significantly altering the electron
fraction and entropy, thereby determining the strength and location of the
initial supernova shock, as well as the entropy and electron fraction
profiles throughout the core.  As a result, improvements in the treatment
of electron capture alter the initial conditions for the entire postbounce
evolution of the supernova.

Calculation of the rate of electron capture on heavy nuclei in the
collapsing core requires two components: the appropriate electron capture
reaction rates and knowledge of the nuclear composition.  The inclusion of
electron capture within a multigroup neutrino transport simulation adds an
additional requirement: information about the spectra of emitted neutrinos.
 Unlike stellar evolution and supernova nucleosynthesis simulations,
wherein the nuclear composition is tracked in detail via a reaction network
\cite{Woos86,HiTh99a}, in simulations of the supernova mechanism, the
composition in the iron core is calculated by the equation of state
assuming nuclear statistical equilibrium (NSE).  Typically, the information
on the nuclear composition provided by the equation of state is limited to
the mass fractions of free neutrons and protons, $\alpha$-particles, and
the sum of all heavy nuclei, as well as the identity of an average heavy
nucleus, calculated in the liquid drop framework \cite{LaSw91}.  In most
recent supernova simulations (see, \eg, \cite{LMMH01,RaJa00,FrWa02}), the
treatment introduced by Bruenn \cite{Brue85} is used.  This prescription
treats electron capture on heavy nuclei through a generic $0f_{7/2}
\rightarrow 0f_{5/2}$ Gamow-Teller resonance \citep{BBAL79} in the average heavy nucleus
identified by the equation of state.  Because this treatment does not
include additional Gamow-Teller transitions, forbidden transitions, or
thermal unblocking (see \cite{LMSD03}), electron capture on heavy nuclei
ceases when the neutron number of the average nucleus exceeds 40.  As a
result, electron capture on protons dominates the later phases of collapse.

As a major advance over this simple treatment of nuclear electron capture,
we have developed a treatment based on recent shell model electron capture
rates from Langanke \& Mart\'inez Pinedo (LMP) \cite{LaMa00} for $45<
A\leq65$ and 80 reaction rates from a hybrid shell model-RPA calculation
\cite{LaKD01,LMSD03} for a sample of nuclei with $A=66-112$.  For the
distribution of emitted neutrinos, we use the approximation described by
Langanke \etal\ \cite{LaMS01}.  To calculate the needed abundances of the
heavy nuclei, a Saha-like NSE is assumed, including Coulomb corrections to
the nuclear binding energy \cite{HixPhD,BrGa99}, but neglecting the effects
of degenerate nucleons \cite{ElHi80}.  This NSE treatment has been used in
prior investigations of electron capture in thermonuclear supernovae
\cite{BDHI00}.  We use the combined set of LMP and hybrid model rates to
calculate an average neutrino emissivity per heavy nucleus.  The full
neutrino emissivity is then the product of this average and the number
density of heavy nuclei calculated by the equation of state.  With the
limited coverage of rates for $A>65$, this approach provides the most
reasonable estimate of what the total electron capture would be if rates
for all nuclei were available.  This averaging approach also makes the rate
of electron capture consistent with the composition returned by the
equation of state, while minimizing the impact of the limitations of our
NSE treatment.   A more detailed description of our method, including tests
of some assumptions made, will be presented in a forthcoming article
\cite{HMMS03}.

Simulations of the collapse, bounce, and post-bounce evolution of a 15
\msun\ model \cite{HLMW01} were carried out using the fiducial Bruenn
prescription for electron capture on nuclei and our LMP+hybrid treatment
with our fully general relativistic, spherically symmetric AGILE-BOLTZTRAN
code.  In these simulations, it is employed using the equation of state of Lattimer and Swesty \cite{LaSw91}, 6-point Gaussian quadrature to discretize the neutrino angular distributions, and 12 energy groups to discretize the neutrino spectra between 3 to 300 MeV.

\begin{figure}
  \includegraphics[width=.4\textwidth]{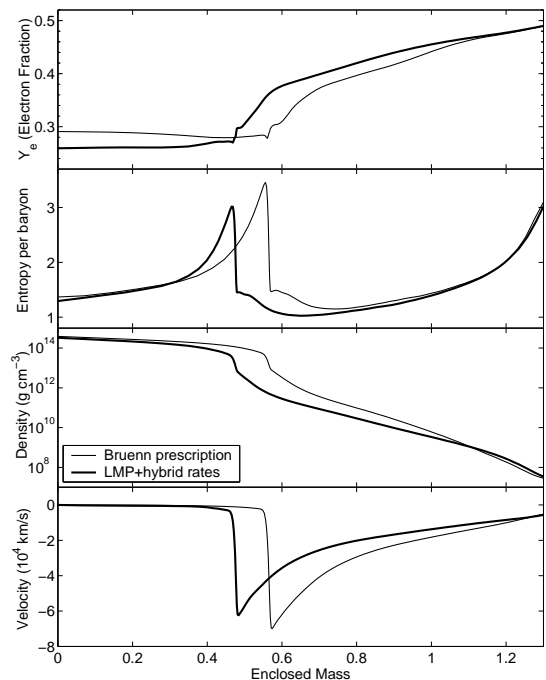}
  \caption{The electron fraction, entropy, density, and velocity as  
  functions of the enclosed mass at the beginning of bounce for a 15 \msun\ 
  model. The thin line is a simulation using the Bruenn parameterization 
  while the thick line is for a simulation using the LMP and hybrid reaction 
  rate sets.
  \label{fig:bounce}}
\end{figure}

Our improved treatment of nuclear electron capture has two competing
effects.  In lower density regions, where the average nucleus is well below
the $N=40$ cutoff of electron capture on heavy nuclei, the Bruenn
parameterization results in more electron capture than the LMP+hybrid
treatment.  This is similar to the reduction in the amount of electron
capture seen in stellar evolution models \cite{HLMW01} and thermonuclear
supernova \cite{BDHI00} models when earlier parameterized rates
\cite{FuFN85} are replaced by shell model calculations.  In denser regions,
the continuation of electron capture on heavy nuclei alongside electron
capture on protons results in more electron capture in the LMP+hybrid case.
The results of these competing effects can be seen in the upper pane of
Figure~\ref{fig:bounce}, which shows the distributions of electron
fraction, entropy, density, and velocity throughout the core at bounce
(maximum central density).

In addition to the marked reduction $(\sim 10\%)$ in the electron fraction
in the interior of the PNS, the improved treatment of electron capture also
results in an $\sim 20\%$ reduction in the mass of the homologous core,
consistent with the analysis that the size of the homologous core is
proportional to the square of the mean trapped lepton fraction
$<{Y_{l}}^{2}>$ at core bounce \cite{Yahi83}.  At bounce, this change in
the homologous core manifests itself as a reduction in the mass interior to
the formation of the shock from 0.57 \msun\ in the fiducial case to 0.48
\msun\ in the LMP+hybrid case, as is evident in the lower pane of
Figure~\ref{fig:bounce}. In the LMP+hybrid case there is also an $\sim
15\%$ reduction in the central density and an $\sim 5\%$ reduction in the
central entropy at bounce, as well as an $\sim 15\%$ smaller velocity
difference across the shock.

\begin{figure}
  \includegraphics[height=3in,width=.4\textwidth]{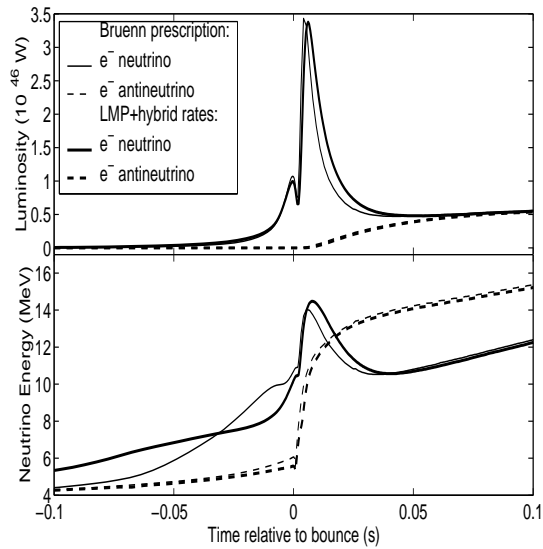}
  \caption{The neutrino luminosity and root-mean-square energy (at 500 km)
  as a function of time from bounce for a 15 \msun\ model. The thin lines
  show this evolution for a simulation using the Bruenn parameterization
  while the thick lines show this evolution for a simulation using the
  LMP+hybrid rates.  The solid lines correspond to electron type neutrinos,
  the dashed lines correspond to electron type antineutrinos.
  \label{fig:nutime}}
\end{figure}

If the principle effect of our improvement in the treatment of electron
capture on nuclei were to launch a weaker shock with more of the iron core
overlying it, this improvement would make a successful explosion more
difficult.  However, these improvements in nuclear electron capture also
alter the behavior of the outer layers which play an important role in the
ultimate fate of the shock.  The lesser neutronization in the outer layers
slows the collapse of these layers, which further diminishes the growth of
the electron capture rate by reducing the rate at which the density
increases.  These changes are clearly apparent in regions above the shocks
in Figure~\ref{fig:bounce}, for example, reductions of a factor of 5 in
density and 40\% in velocity are evident in the vicinity of 0.8 \msun. 
Such changes reduce the ram pressure opposing the shock, easing its outward
progress.  In these spherically symmetric models, these improvements allow
the shock in the LMP+hybrid case to reach 168 km, relative to 166 km in the
fiducial case, in spite of the greater mass overlying the shock when it was
launched and the greater loss of energy to the neutrino burst (see
Figure~\ref{fig:nutime}).

As mentioned earlier, changes in the electron capture rates also lead to
changes in the core fluid gradients that may, in turn, drive fluid
instabilities that are potentially important to the supernova mechanism. 
Within the inner 50 km, the entropy and lepton fraction gradients found in
the LMP+hybrid model are considerably different from those found in the
fiducial model. Consequently, the more accurate treatment of electron
capture may significantly alter the location, extent, and strength of
proto-neutron star convection, or other potential fluid instabilities, in
the core.  This provides an excellent example of the coupling of convective
behavior to radiative and nuclear processes and must be further
investigated in the context of future multidimensional models.

Figure~\ref{fig:nutime} shows the luminosity and mean energy of the emitted
electron neutrinos and antineutrinos between 100 milliseconds before bounce
and 100 milliseconds after bounce.  Clearly evident in the luminosity is a
slight delay (2 ms) in the prominent ``breakout'' burst caused by the
deeper launch of the shock in the LMP+hybrid case.  Over the first 50
milliseconds after bounce, the LMP+hybrid model emits $\sim 15\%$ more
energy than the fiducial model, with a slightly lower luminosity at later
times.  This is largely the result of differences in the mean electron
neutrino energy, which is as much as 1 MeV higher over the first 50
milliseconds in the LMP+hybrid case, but lower thereafter.  This results
from the neutrinospheres in the LMP+hybrid model occurring in deeper,
hotter layers for the first 50 milliseconds, but cooler layers at later
times.

The differences in the neutrino spectrum during collapse, when electron
capture on nuclei dominates, are larger than those described after bounce. 
For low densities, where capture on nuclei dominates in the Bruenn
prescription as well, the approximate reaction Q-value derived from the
free neutron and proton chemical potentials dramatically underestimates the
Q value, resulting in a much lower mean neutrino energy.  As captures on
protons begin to compete with captures on nuclei in the Bruenn
prescription, the mean neutrino energy grows rapidly because of the higher
Q value for capture on protons.  It exceeds that found in our LMP+hybrid
model by as much as 2 MeV in the 30 milliseconds just before bounce.  This
latter effect was anticipated by Langanke \etal \cite{LMSD03}, who also
demonstrated that nuclear electron capture should dominate that on protons
because the much larger number of heavy nuclei more than compensates for
the larger capture rate on free protons, at least for the conditions found
in models resulting from the Bruenn prescription.  These self consistent
models unequivocally show that this is correct.  At the onset of collapse,
there are roughly 1000 heavy nuclei per proton in the inner layers of the
core.  In our fiducial model this ratio declines rapidly, reaching values
less than 100 by the time the central density is $10^{11} \gcc$ and roughly
10 by the time the central density is $10^{13} \gcc$.  In the LMP+hybrid
model, the reduced electron fraction and entropy keep this ratio near 1000
until the central density exceeds $10^{12} \gcc$, reaching 50 around a
central density of  $10^{13} \gcc$.  As a result, in the regime where
$Y_{e}$ experiences the largest changes ($10^{11-13} \gcc$), the dominance
of heavy nuclei is increased by a factor of 5-30, cementing the dominance
of nuclear electron capture.


To implement these simulations, we have made approximations to all three
components of the calculation of electron capture in core collapse
supernovae.  Each of these requires further improvement.  Removing the
averaging of electron capture rates requires better coverage of electron
capture on nuclei in the region $65<A<120$, by hybrid and other approximate
methods \cite{PrFu03} in the near term, but ultimately by shell model
calculations vetted by experimental determinations of Gamow-Teller strength
distributions.  These reaction rates must cover the full thermodynamic
range of interest in supernovae (temperatures of 1-100 \gk\ and densities
from $10^5 - 10^{14} \gcc$) and must also address the need for the emitted
neutrino spectra.  Detailed tracking of the nuclear composition is also
necessary, in a form that retains the consistent transition to nuclear
matter afforded by current schemes \cite{LaSw91} while allowing for
accurate calculation of the rate of electron capture on heavy nuclei and,
ultimately, for detailed nucleosynthesis.

We have demonstrated that supernova simulations with a modern treatment of
electron capture differ significantly from previous models, which employed
more parameterized treatments.  Though this improved model still fails to
produce an explosion in the spherically symmetric case, the differences are
quite striking.  The initial mass behind the shock when it is launched is
reduced by 20\%, with significantly $(\sim10\%)$ lower central densities,
entropies, and electron fractions in this region.  Over the first 50 ms
after bounce, the neutrino luminosity is boosted by $\sim15\%$ with the
mean electron neutrino energy increased by $\sim 1 $ MeV.  In spite of an
initially weaker and deeper shock and larger neutrino energy loss, reduced
electron capture in the outer layers slows their collapse, allowing the
shock to reach a maximum radius that is slightly larger.  Furthermore the
lepton and entropy gradients in the core differ significantly.  Because
these gradients drive PNS convection and other potential instabilities in
the core, the location and strength of such instabilities may be
significantly different than heretofore thought.

\begin{acknowledgments}
The authors acknowledge helpful conversations with J. Beacom and H.-T.
Janka. The work has been partly supported by NASA under contract NAG5-8405,
by the National Science Foundation under contract AST-9877130, by the
Department of Energy, through the PECASE and Scientic Discovery through
Advanced Computing Programs, by funds from the Joint Institute for Heavy
Ion Research and by the Danish Research Council.  Oak Ridge National
Laboratory is managed by UT-Battelle, LLC, for the U.S. Department of
Energy under contract DE-AC05-00OR22725.
\end{acknowledgments}

\end{document}